# A Novel Non-Volatile Inverter-based CiM: Continuous Sign Weight Transition and Low Power on-Chip Training

Dong Zhang[+], Yuye Kang[+], Gan Liu[+], Zuopu Zhou, Kaizhen Han, Chen Sun, Leming Jiao, Xiaolin Wang, Yue Chen, Qiwen Kong, Zijie Zheng, Long Liu, Xiao Gong*

National University of Singapore, Singapore;

[+]: Equal Contribution;

*Email: elegong@nus.edu.sg;

**Abstract**

In this work, we report a novel design, one-transistor-one-inverter (1T1I), to satisfy high speed and low power on-chip training requirements. By leveraging doped $HfO_2$ with ferroelectricity, a non-volatile inverter is successfully demonstrated, enabling desired continuous weight transition between negative and positive via the programmable threshold voltage ($V_{TH}$) of ferroelectric field-effect transistors (FeFETs). Compared with commonly used designs with the similar function, 1T1I uniquely achieves pure on-chip-based weight transition at an optimized working current without relying on assistance from off-chip calculation units for signed-weight comparison, facilitating high-speed training at low power consumption. Further improvements in linearity and training speed can be obtained via a two-transistor-one-inverter (2T1I) design. Overall, focusing on energy and time efficiencies, this work provides a valuable design strategy for future FeFET-based computing-in-memory (CiM).

# Introduction

Artificial intelligence (AI) has been widely employed as an indispensable part of nowadays advanced intelligent tasks and delivering significant improvements over a wide range of applications, from image processing to autonomous driving [1-2]. For those tasks, heavy workload commonly exists, and at the same time, the total data capacity keeps rising explosively. Thus, the high-speed and low-power graphics processing unit (GPU) is called for [3]. However, the further improvements in performance and energy efficiency of today's GPU design will be ultimately limited by the traditional Von-Neumann architecture [4-7], as most of the time and energy are dissipated during data transmission, in which frequently shuttling back and forth between the processing unit and the memory is inevitable [8-9]. According to earlier studies, the energy consumption for memory access is two orders higher than that for calculation [10]. To continuously push the frontier of AI, computing in memory (CiM) has been proposed as a promising architecture for the next generation to minimize the cost of data transfer. In recent years, various memory types, such as resistive random access memory (RRAM), Flash, and phase change memory (PCM), have been comprehensively studied and demonstrated [11-15]. For instance, Yao et al. have demonstrated a fully hardware-implemented memristor convolutional neural network [16], and Feng et al. have proposed a general-purpose CiM scheme based on 55nm NOR Flash technology [17]. However, those designs suffer from various inherent drawbacks, such as high working current of RRAM and slow operation of NOR Flash [18-20]. Meanwhile, ferroelectric field-effect transistor (FeFET) is becoming attractive due to its high on-off ratio, decent endurance, as well as symmetrical linearity, and has been employed in various CiM schemes based on on-chip and off-chip training [21-24].

On-chip training stands out among the two schemes for alleviating the impacts of stringent area constraints, such as circuit-parameter fluctuation, which significantly affects training accuracy [25-26]. In previous reports of on-chip training implementations, complementary-cell design, like two-transistor-two-FeFET (2T2F), is commonly adopted for mitigating the IR drop effect to boost calculation accuracy, where the output current can be canceled out automatically by two opposite currents [27-28]. Nevertheless, the implementation of continuous sign weight transition in it is challenging when performing weight updating, since two gates of 2T2F are programmed separately, as shown in **Fig. 1 (a)**. To overcome this shortcoming, additional off-chip storage of the weight data mapped in arrays and extra off-chip calculations based on Von Neumann architecture for signed-weight comparison are required, leading to speed and energy consumption concerns, as shown in **Fig. 1 (b)**. To enable a smooth transition between negative and positive weight, an improved version, 2T2F with weight bias (2T2F-WB), is designed by pairing a weight-fixed cell with a weight-tunable cell. However, it suffers from a much higher working current than conventional 2T2F design when representing same weight, as shown in **Fig. 2 (a)**. Hence, serious problems, including low energy efficiency and IR drop-induced training-accuracy degradation, limit its application potential. To achieve continuous sign weight transition at a well-optimized working current, the unit cell design driven by new mechanisms is urgently demanded.

In this work, a novel design comprising a one-transistor-one-inverter (1T1I) is proposed and the non-volatile inverter is experimentally demonstrated to functionalize a continuous sign weight transition at an optimized working current. The operation principle and the feasibility of the proposed design are validated by simulations based on a dynamic ferroelectric model and experiments. In addition, circuit level figure-of-merits, including energy consumption, IR drop, and training accuracy, are comprehensively evaluated and benchmarked. Finally, an improved design based on hybrid precision, two-transistor-one-inverter (2T1I), has been further proposed for future neuron network training to fulfill the requirements of decent linearity, high speed, and better accuracy.

**Principle and Simulation**

**Fig. 2 (c)** elaborates the operation principle of sign weight transition between negative and positive realized in the proposed non-volatile inverter, which comprises an n-FeFET and a p-FeFET. To simplify the schematic, the selective transistor employed for sneak-path current elimination is not depicted here. Different from orthodox inverter design, for power supply, the source terminal of p-FeFET is connected to a positive bias, and that of n-FeFET is connected to a negative bias. The detailed sign weight transition mechanism is described as follows: by programming the current direction at the output node, the cell contributes to the charge-integral circuit in peripherals positively (current out) or negatively (current in), deciding the sign of the weight to be + or -. For example, when the p-FeFET is on, and the n-FeFET is off after a negative pulse-programming, the sign of weight is positive, and vice versa. In addition to the sign transition, the desired multi-state weight updating can be achieved by a continuous conductance adjustment enabled by polarization switching induced threshold voltage ($V_{TH}$) shift. Regarding the programming scheme, it should be noted that our design enjoys a significant benefit compared with the conventional 2T2F design: instead of involving two separate programmable gates, the two FeFETs in our 1T1I structure share a common gate. Via inputting positive or negative pulses to the common gate, the $V_{TH}$ of both FeFETs shift simultaneously at the same amplitude and direction. In this way, a continuous sign weight transition with multiple states is realized without relying on any off-chip calculation assistance.

Based on our released dynamic ferroelectric model [29], simulations have been performed based on a non-volatile inverter structure to verify the proposed theory. For both FeFETs, the channel length is set as 100 nm, and the ferroelectric layer thickness is set as 10 nm. Transfer curves of them with prominent hysteresis are shown in **Fig. 3 (a)** and **(b)**, with memory windows of around 1.5 V, which are sufficient to fulfill the requirements of various CiM applications. Besides, as shown in the schematic, the hysteresis in p-FeFET is clockwise, while for n-

FeFET, it is anticlockwise. The reason is that the types of majority carriers in n-FeFET and a p-FeFET are different: electrons are dominant in n-FeFET, while holes play a substantial role in p-FeFET.

The discriminable minimum current resolution (MCR) at switching point under a given voltage step, 10 mV in this work, is determined by the position of crosspoint. Therefore, via playing $V_{TH}$-engineering on FeFETs, the operation mode of the non-volatile inverter can be designed to be high-speed (high working current) and low-power (low working current) to satisfy different requirements under various application scenarios. The current switching characteristics around the crosspoint at two different resolutions are shown in **Fig. 3 (c)** (resolution ~1 µA) and **(d)** (resolution ~0.01 µA). Output current ($I_{OUT}$) as a function of $V_{IN}$ and the voltage transfer curves of the non-volatile inverter are further simulated and plotted in **Fig. 3 (e)** and **(f)**, respectively. A clear curve shift caused by the polarization of ferroelectricity can be observed, supporting the feasibility of the proposed scheme.

## Experiment and Measurement

### A. Device Fabrication

The devices are fabricated on a silicon-on-insulator (SOI) wafer with ~220 nm Si layer as the first layer and ~3 µm buried $SiO_2$ on a Si substrate. A cyclic surface cleaning is performed prior to the fabrication using diluted hydro-fluoride (DHF) and deionized water (DI water). The channel of Si p-FeFET is then defined by laser writer using AZ1512 as the photoresist. This is followed by boron implantation at source/drain (S/D) regions with a dose of $1 \times 10^{15}$ cm$^{-2}$ and an energy of 25 keV. After the dopant activation at 1000 °C for 5 s in $N_2$ ambient using rapid thermal processing (RTP), the gate stack is formed with 10 nm doped $HfO_2$ and 30 nm W through atomic layer deposition (ALD) and sputtering. The sample is then annealed at 450 °C for 30 s in nitrogen ambient to crystallize the doped $HfO_2$ film using RTP. Finally, titanium/gold (Ti/Au) bilayer stack is deposited by e-beam evaporator and lifted off in acetone to form the S/D contact. For Si n-FeFET, the gate stack comprising doped-$HfO_2$ and W metal gate is formed before the phosphorus implantation at S/D regions. The doped $HfO_2$ crystallization and phosphorus activation are performed together at 600 °C for 30 s in RTP. Meanwhile, metal-ferroelectric-metal (MFM) capacitors with ~10 nm doped $HfO_2$ film are also fabricated to confirm the ferroelectricity of the FE film applied in the transistors. The top electrode and bottom electrode of the MFM capacitor use the same metal of W.

**B. Electrical Characterization**

The fabricated MFM capacitors are firstly characterized using the FE analyzer and B1500A semiconductor analyzer. The ferroelectricity of the FE film is confirmed by the polarization-voltage (P-V) loop, as described in **Fig. 4 (a)**. With an enlarged voltage range, remnant polarization and current peaks increase rapidly. The remnant polarization could achieve ~20 μC/cm$^2$ at a voltage bias of 4 V.

After the characterization of the FE film, the electrical characterizations of Si FeFETs and inverters are carried out using a Keithley 4200 semiconductor analyzer at room temperature. **Fig. 4 (b)** and **(c)** show the transfer curves of Si p-FeFET with the channel length ($L_{CH}$) of 50 μm and Si n-FeFET with $L_{CH}$ of 8 μm at the low drain voltage ($V_{DS}$) of ±0.05 V, respectively. The drain current is normalized by the channel width of 50 μm. Overlapping area between gate and S/D is applied to both FeFETs to facilitate polarization switching through gate-induced-drain-leakage (GIDL) [30]. Slight GIDL improves erasing characteristic due to enough carriers supplied from heavily doped S/D regions at the inversion state. The p-FeFET has a memory window of 0.51 V, and the n-FeFET shows a memory window of 0.84 V. Some electrical characteristics of our p-FeFET, like off-state current ($I_{OFF}$), on-current/off-current ratio ($I_{ON}/I_{OFF}$), and subthreshold swing (*SS*), are not satisfactory due to the lack of surface treatment and device-to-device isolation. Optimized surface treatment and device-to-device isolation can be applied to further enhance the device performance in future study. Meanwhile, |$V_{TH}$| and on-state current ($I_{ON}$) of p-FeFET and n-FeFET can be further adjusted for better inverter characteristics [31]. **Fig. 4 (d)** and **(e)** depict the retention characteristics of Si p-FeFET and n-FeFET after fully programming and erasing. The drain current measured at the gate voltage ($V_{GS}$) of -0.5 V and |$V_{DS}$| of 0.05 V shows negligible degradation. Hence, after programming and erasing, the states stored in both p-FeFET and n-FeFET can be maintained to beyond 10 years by extrapolating the measured curves.

The inverter characteristics are then measured by connecting separate p-FeFET and n-FeFET. The inset of **Fig. 4 (f)** depicts the equivalent circuit schematic of the connection between Si p-FeFET and n-FeFET, and the main part plots output voltage ($V_{OUT}$) versus $V_{IN}$ curves of the inverter. By virtue of the polarization switching, the inverter is successfully programmed to two different states by a positive $V_{IN}$ (POS PROG) or a negative $V_{IN}$ (NEG PROG) with a large amplitude. A distinct shift between the voltage transfer curves can be observed. When it comes to low power consumption, it should also be noted that our inverter works at low $\pm V_{DD}$ of $\pm 0.5$ V, showing compatibility with energy-saving systems operated at low supply voltage. The voltage gain ($\Delta V_{OUT}/\Delta V_{IN}$) is further extracted as a function of $V_{IN}$, as shown in **Fig. 4 (g)**. A peak voltage gain of 7 V/V is obtained with POS PROG as the initial state. However, the peak voltage gain degrades to 5 V/V when NEG PROG is the initial state. The difference between the voltage gains obtained from the different states indicates that the transconductance and output resistance of the FeFETs get varied from state to state. Hence, as compared with MOSFET-based inverter, whose performance is fixed after fabrication, the performance of FeFET-based inverter can be tuned after the devices have been finalized. This might be helpful in providing a more flexible circuit design scheme beyond this work. As for the most attractive feature of the proposed non-volatile inverter, the transition between the negative output current and positive output current is achieved experimentally as well. **Fig. 4 (h)** exhibits $|I_{OUT}|$ versus $V_{IN}$. Valley shape can be clearly observed for both curves.

The lowest point of the valley represents a current transition between negative and positive. The inset figure illustrates the direction of the current flow through p-FeFET and n-FeFET. A notable shift of the transition point (the lowest point of the valley) is observed by plotting $|I_{OUT}|$ in the log scale. **Fig. 4 (i)** further zoom-in the $V_{IN}$ to the range of -0.6 V ~ -0.7 V, and $I_{OUT}$ is plotted on the linear scale. Within the range, the output current can be changed from p-FeFET dominated (purple line) to n-FeFET dominated (green line) by programming their conductance. Apparently, with the assistance of ferroelectricity, $I_{OUT}$ can be switched between two opposite directions with comparable amplitude at a fixed $V_{IN}$, indicating the realization of a transition between positive and negative weight.

## Circuit Performance Evaluation

Including the subarray and peripheral circuits, **Fig. 5 (a)** shows the CiM layout of the processing element (PE) that supports both forward and backward propagations in on-chip training process. The switch matrix and the multiplexer composed of transmission gates (TG) are introduced in this design to simplify peripheral circuits for better energy efficiency and space-saving. Some other key components, such as subtractors, shifters, and adders, are not shown in the schematic. However, the evaluations in this work carefully consider their influences to get convincing results, and the functions of the modules are elaborated as follows: subtractors are required during the back propagation process to subtract the currents carrying negative weights from those carrying positive weights, since they cannot be cancelled out locally due to their same flowing direction [32]; the shifters and adders function when considering the weights of input vectors. After going through the processing circuit in PE, the output data will be transferred to the function unit waiting for further instructions.

**Fig. 5 (b)** shows the schematic of internal current flows in the synaptic array to illustrate the source of IR drop. In the scenario of a design for CiM applications, due to the relatively small scale of the memory array, IR drop is mainly contributed by the non-ideal series resistance of TG instead of the wire resistance. To be more specific, the on-state resistance of the TG in CiM circuit is normally at several kilo-Ohm level, while the wire resistance for a $128 \times 128$ array is around one hundred Ohm [33]. Intuitively speaking, enlarging channel width to suppress the non-ideal resistance and IR drop compensation via supply voltage adjustment could be two feasible solutions. However, sacrificing footprint is undesired for the realization of a compact system. In addition, the IR drop is determined by the weights stored in the array and the vector information of input, leading to an impossible compensation by simply adjusting supply voltage due to its randomness. Therefore, the optimization of the working current is much sought after. For mapping 5 bits/unit ranging from -15 to 15, the 1T1I design consumes a much smaller unit current as compared with that of 2T2W-WB scheme, and the highest current ratio is up to 31 times, indicating a decent IR drop suppressing capability of the 1T1I design. To further quantify the benefit, based on unit current analysis, the investigation is performed on the normalized average IR drop in back propagation, which is much severer than that in forward propagation due to higher current crowding at the input terminal. The evaluation is conducted in a 128×128 array, in which the $R_{ON}$ of the transmission gate is set as 6 kΩ, and the conductance range of memory cell is set from $-1.5 \times 10^{-7}$ to $1.5 \times 10^{-7}$ S, corresponding to the weight range as mentioned above. As expected, 1T1I has a much lower average IR drop as compared with that of 2T2F-WB observed in **Fig. 5 (c)**: the IR drop is 0.05 for 1T1I and is 0.23 for 2T2F-WB. The reduction is extracted to be ~78%.

Training-accuracy simulation is further performed to evaluate the influence of IR drop. MNIST handwritten data and 100 training epochs are adopted in the simulation. The detailed training process is elaborated here: first of all, the input data of MNIST is converted into binary voltage pulses, and the weight values in convolution and fully connected layers are transformed into hexadecimal weights and mapped in arrays in the form of conductance; subsequently, the vector-matrix multiplication process based on Ohm's law and Kirchhoff's current law is performed [6]. As for other processes, like pooling and accumulation considering different weights, are realized in the function units on the CiM chip [34]. The result in **Fig. 5(d)** shows that by taking advantage of the well-controlled IR drop, 1T1I-based CiM possesses a training accuracy of over 90%, which approximates the software accuracy. However, the accuracy of 2T2F-WB-based one gets degraded to ~80% due to the considerable IR drop.

When it comes to energy consumption at the circuit level, as shown in **Fig. 5**

**(e)**, a benefit of 46% reduction can be obtained when both the array and the peripherals are included in consideration, where the evaluation is conducted based on the publicly available software: NeuroSim [35-36]. The value of energy consumption is obtained by averaging the energy consumed by one forward and one backward propagation. In addition, to mimic the real situation, instead of using the mean-weight value, the weights mapped in cells are extracted directly from the neuro networks. The breakdown of the total energy consumption is listed in **Fig. 5 (f)** to get a deep insight into the improvement gained from 1T1I structure. As shown, for ADC and subtractors, a distinct gap in energy consumption can be observed due to the difference in current magnitude in the two designs, where larger current needs higher precision bit of peripherals to guarantee result accuracy. In contrast, for the shifters and adders, the energy consumption in both scenarios is same. The reason is that the data flow to be processed by them is identical after the subtracting process. In addition, note that the profit is mainly contributed by saving the energy consumption of array: 2T2F-WB scheme shows a larger energy budget due to the high working current dilemma.

In **Fig. 5 (g),** the mechanism of 2T1I design based on the hybrid precision design [37] is illustrated. As to 2T1I, the conductance of the unit cell is controlled by the threshold voltage of the inverter and the voltage at the gate node ($V_{GN}$) simultaneously. Thus, weight updating can be performed by tuning $V_{GN}$ rather than programming the threshold voltage of it, which is time and energy efficient. The adjustment of $V_{GN}$ is achieved by charging and discharging processes with the assistance of two independent control gates: positive pulses are applied at the above terminal for charging, and negative pulses are applied at the bottom terminal for discharging. Thanks to the ingenious design, identical pulses can be utilized for conductance tuning without sacrificing linearity. Moreover, note that two weights (positive and negative) are stored in the 2T1I scheme, while the weight storage in 1T1F is single. Therefore, there is no additional area sacrifice for mapping same-weight matrices in the proposed design.

**Fig. 5 (h)** benchmarks 1T1I design proposed in this work with other widely applied unit devices in nowadays on-chip training implementations: traditional 2T2F and 2T2F-WB. As shown in comparison, 1T1I structure can smoothly realize the continuous sign weight transition between negative and positive, which accelerates the writing speed and reduces the writing energy. In addition, enjoying the advantages of a well-controlled IR drop and optimized current, the 1T1I scheme shows various strengths, including low power consumption and high training accuracy. With all its superior characteristics, the 1T1I design can be appointed to serve as a mainstream structure for on-chip training based CiM.

## Conclusion:

Enabled by a pure on-chip-based sign weight transition at optimized working current, the 1T1I design proposed in this work outperforms the commonly used designs nowadays in many aspects, including but not limited to writing speed, IR drop control, training accuracy, and power consumption. Simulations and experiments performed in this work provide comprehensive studies on the proposed scheme from theory to practice, manifesting the 1T1I design as a possible key enabler in future hardware-implementation of advanced neural networks.

## Acknowledgment:

We acknowledge the funding support from Singapore MOE Tier 1 (R-263-000-C58-133) and MOE Tier 2 (R-263-000-D77-112).

## Author contributions:

X. Gong and D. Zhang proposed this project. X. Gong, Y. Kang, and Z. Zhou conceived and designed the experiments. Y. Kang, Z. Zhou, D. Zhang, and G. Liu performed device fabrication and electrical measurements. D. Zhang carried out the circuit-level simulation. All authors contributed to the discussion and data analysis. K. Han, Y. Kang, D. Zhang, and X. Gong wrote the manuscript.

## Competing Interests：

The authors declare no competing interests.

## Data availability:

The data that support the plots within these paper and other findings of this study are available from the corresponding author on reasonable request.

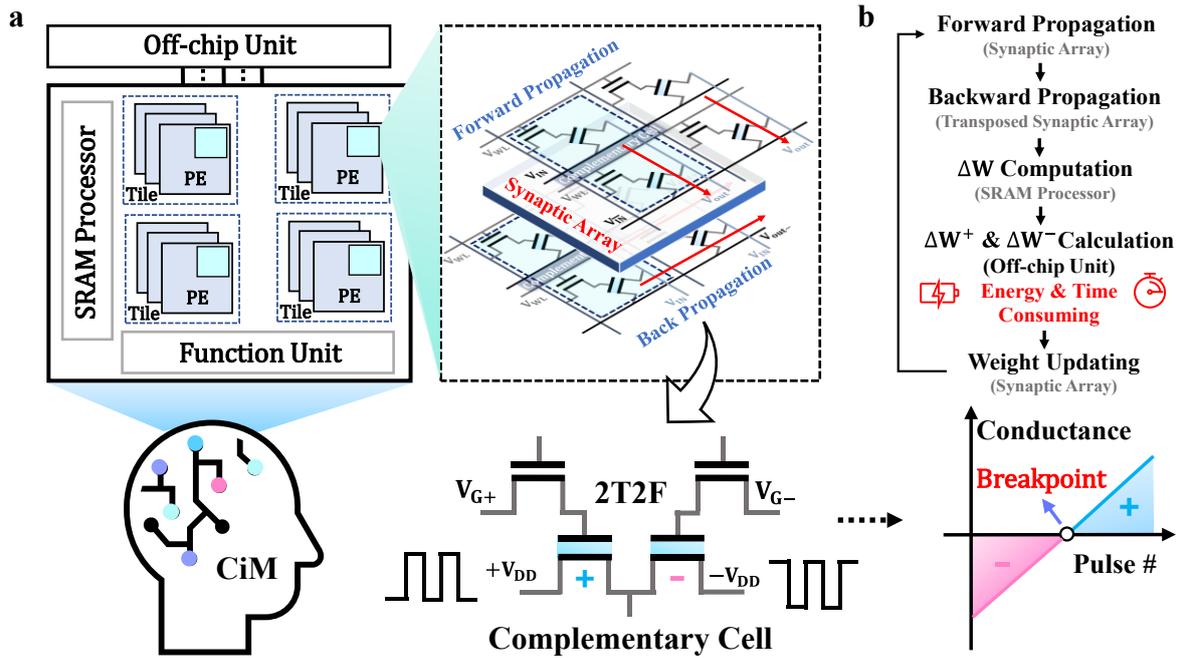

**Fig.1 The schematic of on-chip training implementation with the unit cell of 2T2F. a,** Illustration of on-chip training circuit, in which the processing element is the basic unit to perform vector-matrix multiplication, and the SRAM processer is designed to calculate the weight gradient. Some other processes in on-chip training, such as pooling and activation, are employed in the function unit. The transposable syntactic array containing complementary cells of 2T2F supports both forward and backward propagations. **b,** The flow chart of the on-chip training process. In 2T2F based on-chip training schemes, after the calculation of weight gradient, additional calculations for signed-weight comparison are demanded to avoid programming errors since the weight updating between negative and positive is not continuous due to writing separation.

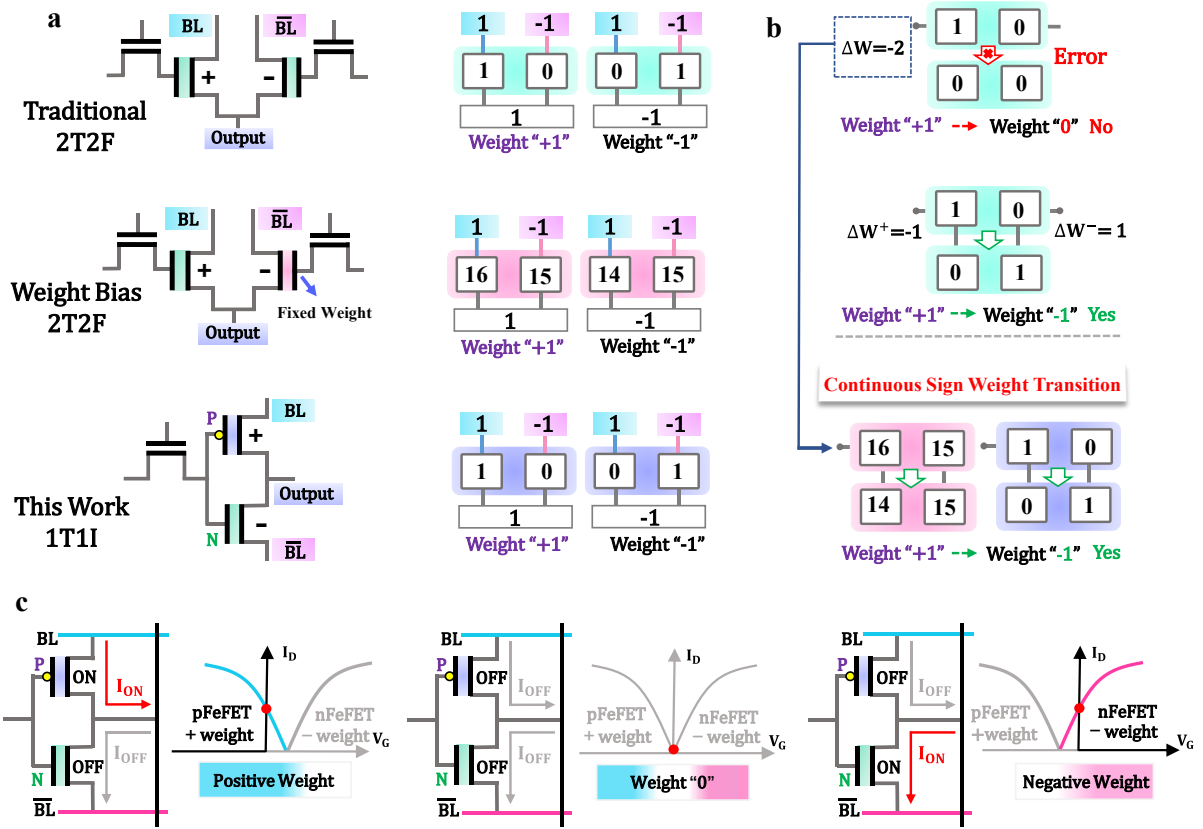

**Fig.2 The schematic and the programming mechanism of various unit cells. a,** The illustration of weight expression for unit cells in on-chip training, such as traditional 2T2F, 2T2F-WB, and 1T1I. The 2T2F-WB and 1T1I structures show an optimized current for representing identical weight, where 5 bits/unit is adopted with the weight range of [-15,15]. **b,** The schematic of weight updating ranges from "1" to "-1". For traditional 2T2F design, extra calculations are involved to eliminate writing errors when the change of weight sign happens. **c,** The illustration of the working principle of 1T1I, in which p-FeFET and n-FeFET store positive and negative weights, respectively. The same direction $V_{TH}$ shift is controlled strictly by the adoption of identical ferroelectric layer thickness. During writing, if positive programming pulses are applied, the $I_D$-$V_G$ curve will shift negatively, making n-FeFET on and p-FeFET off, and vice versa.

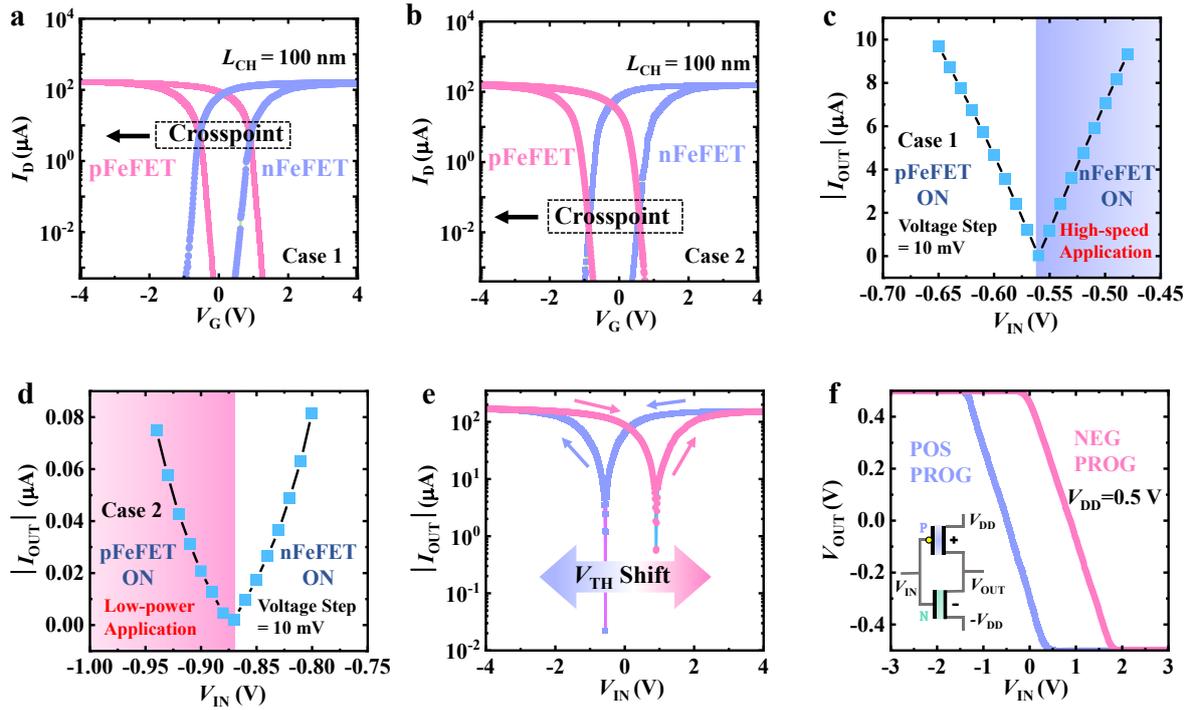

**Fig.3 The simulation results of ferroelectric non-volatile inverter. a, b,** Simulated $I_D$-$V_G$ characteristics for both n-FeFET and p-FeFET in a non-volatile inverter. The hysteresis for n-FeFET is clockwise, while in p-FeFET, it is anticlockwise. The finest current resolution of the non-volatile inverter can be programmed by tuning the crosspoint between the transfer curves. **c, d,** The $I_{OUT}$-$V_{IN}$ curves of the non-volatile inverter with different crosspoints. By proper current resolution engineering, the non-volatile inverter can be switched between the modes of high-speed (high current) **(c)** and low-power (low current) **(d)**. **e,** Simulation result of the $I_{OUT}$-$V_{IN}$ curve of the non-volatile inverter. An obvious shift of the transition point contributed by the polarization of ferroelectric layer can be observed. **f,** Simulation result of the voltage transfer curve of the non-volatile inverter.

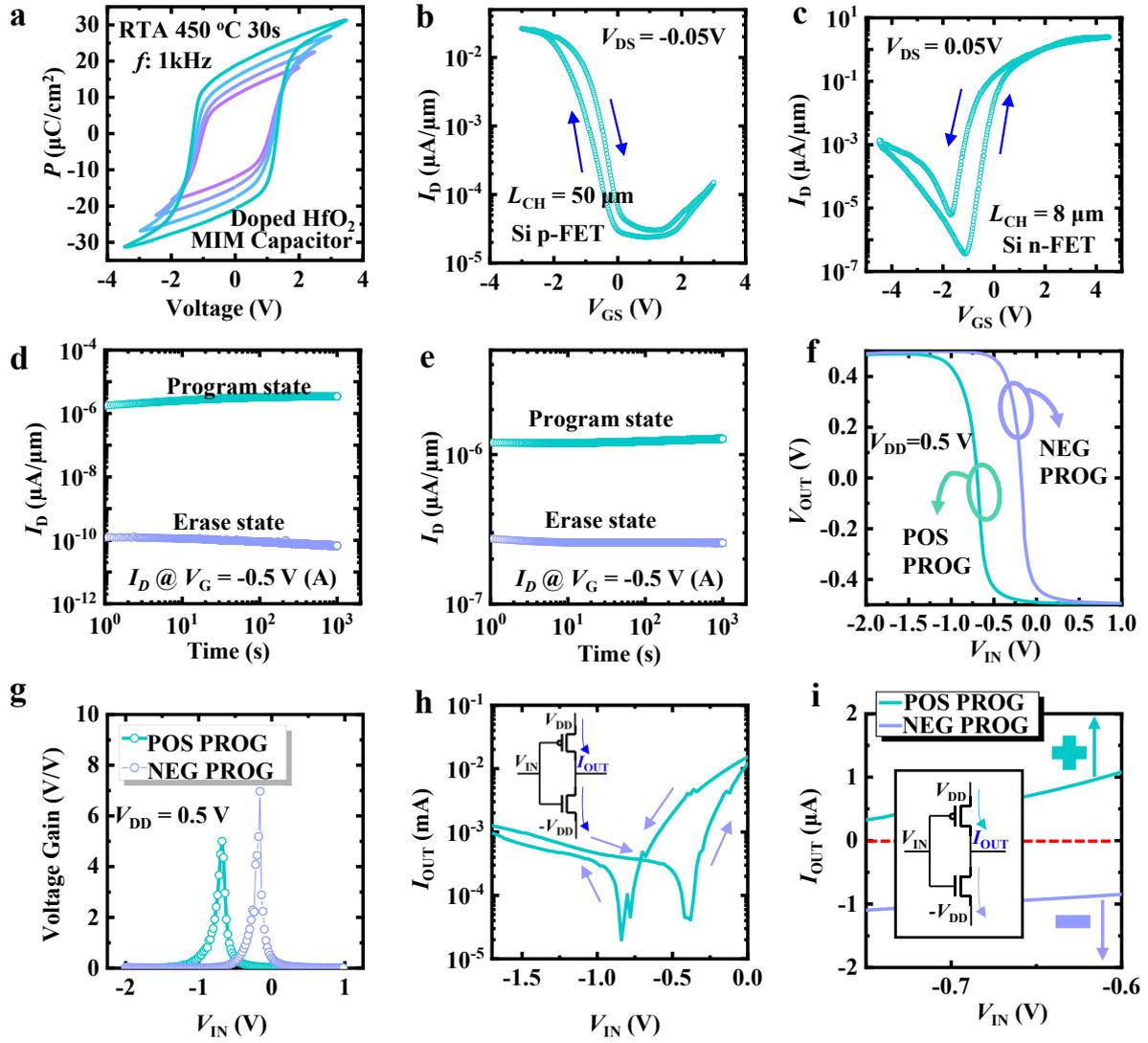

**Fig.4 The experimental demonstration of non-volatile inverter. a,** Polarization-voltage loops of doped $HfO_2$ MFM capacitors. **b, c,** Transfer curves of Si **(b)** p-FeFET with 50 μm $L_{CH}$, and **(c)** n-FeFET with 8 μm $L_{CH}$, at $|V_{DS}|$ of 0.05 V. Overlap structure is applied to facilitate MW. **d, e,** Measured retention characteristics of Si **(d)** p-FeFET and **(e)** n-FeFET after fully programming and erasing. $I_D$ measured at the same $V_G$ of -0.5 V shows negligible degradation. **f,** Voltage transfer curves of the inverter comprising Si p- and n-FeFETs. **g,** The voltage gain versus $V_{IN}$ shows the highest gain of 7 V/V with POS PROG. **h,** Notable $V_{TH}$ shift from $I_{OUT}$ versus $V_{IN}$ plotted in log scale. **i,** The inverter can be programmed between two opposite states at same $V_G$.

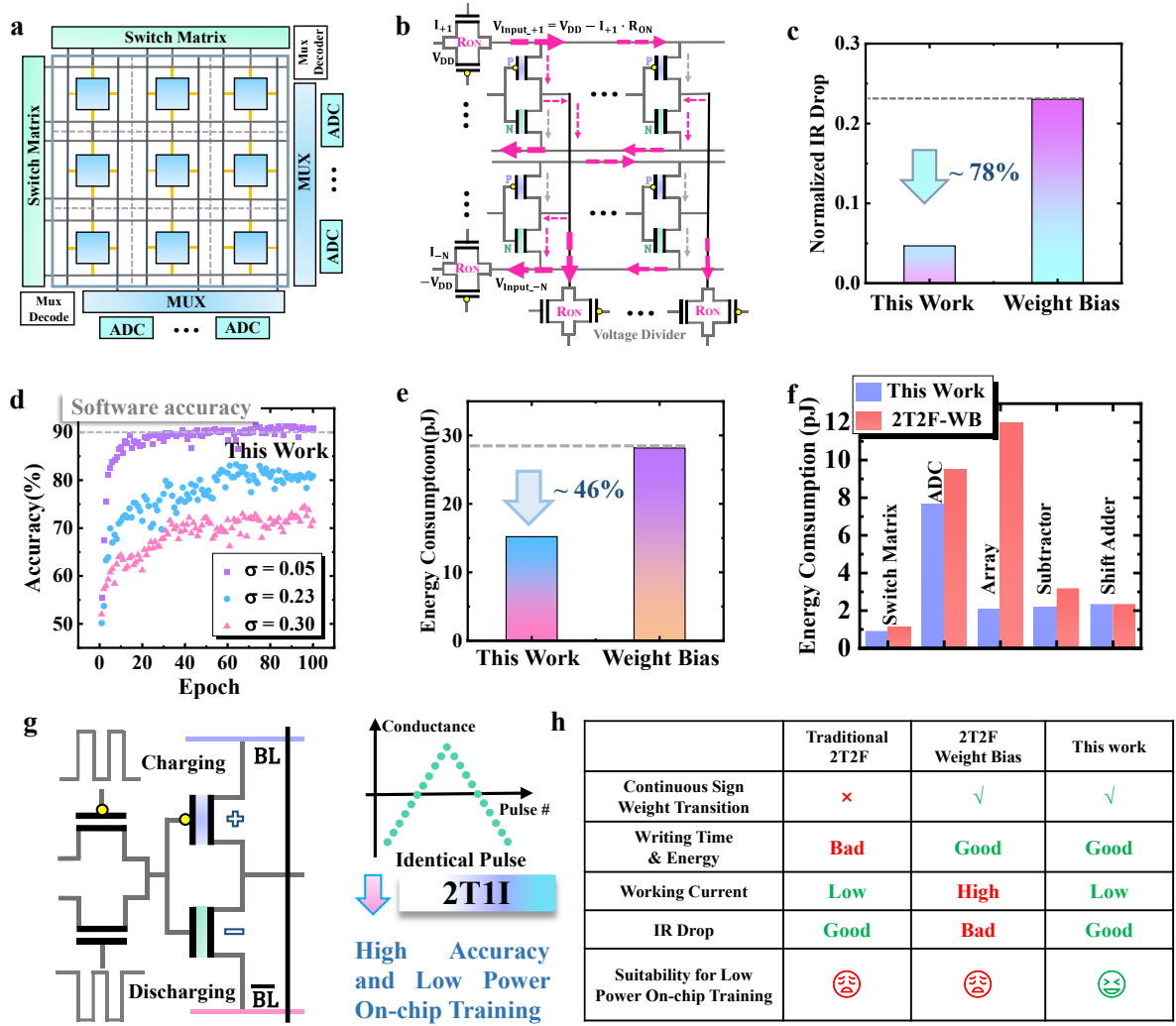

**Fig.5 Circuit evaluation and analysis of 1I1T based CiM. a,** The schematic of the CiM circuit supporting both forward and back propagations of on-chip training. **b,** The illustration of the main derivation of IR drop in subarray: non-ideal resistance of TG, which is at kilo-Ohm level. **c,** The comparison of average IR drop with different unit cells, 5 bits/unit (including sign weight) is adopted. **d,** The comparison of training accuracy under various IR drops. **e,** The evaluation of energy consumption at circuit level, considering peripherals and array. **f,** The energy breakdown of the main part in a synaptic array. **g,** The schematic of the hybrid precision-based design, 2T1I, towards high speed and low-power on-chip training process. **h,** Benchmark table with other unit cell designs supporting on-chip training. 1T1I is more suitable for on-chip training by enjoying the benefits of higher writing speed, lower working current, and smaller IR drop.